\documentclass[12pt]{article}
\usepackage{euler}
\usepackage{amsmath}
\usepackage{amssymb}
\usepackage{amstext}
\usepackage{amscd}
\usepackage{amsfonts}
\DeclareMathAlphabet{\EuFrak}{U}{euf}{m}{n}
\DeclareMathAlphabet{\EuScript}{U}{eus}{m}{n}

\title{{\bf Bosonic String and String Field Theory: a solution 
using Ultradistributions of Exponential Type}
\thanks{\it{This work was partially supported by Consejo
Nacional
de Investigaciones Cient\'{\i}ficas and Comisi\'{o}n de
Investigaciones Cient\'{\i}ficas de la Pcia. de Buenos
Aires;
Argentina.}}}
\author{C.G.Bollini and M.C.Rocca\\
Departamento de F\'{\i}sica, Fac. de Ciencias Exactas,\\
Universidad Nacional de La Plata.\\
C.C. 67 (1900) La Plata. Argentina.}

\date{Ferbruary 1, 2007}

\begin{document}

\maketitle

\vspace{-5mm}

\begin{abstract}

In this paper we show that 
Ultradistributions of Exponential Type
(UET) are appropriate for the description
in a consistent way string and string field theories.
A new Lagrangian for the closed string is obtained
and shown to be equivalent 
to Nambu-Goto's  Lagrangian. 
We also show that the string field 
is a linear superposition of UET of compact support CUET).
We evaluate the propagator  for the string field,
and calculate the convolution of two of them.


\end{abstract}

\newpage

\renewcommand{\theequation}{\arabic{section}.\arabic{equation}}

\section{Introduction}

In a serie of papers \cite{tp1,tp2,tp3,tp4,tp5}
we have shown that Ultradistribution theory of 
Sebastiao e Silva  \cite{tp6,tp7,tp8} permits a significant advance in the treatment 
of quantum field theory. In particular, with the use of the 
convolution of Ultradistributions we have shown that it  is possible
to define a general product of distributions ( a product in a ring
with divisors of zero) taht sheds new ligth on  the question of the divergences
in Quantum Field Theory. Furthermore, Ultradistributions of Exponential Type  
(UET) are  adequates to describe
Gamow States and exponentially increasing fields in Quantun 
Field Theory \cite{tp9,tp10,tp11}.

Ultradistributions also have the
advantage of being representable by means of analytic functions.
So that, in general, they are easier to work with  and,
as we shall see, have interesting properties. One of those properties
is that Schwartz's tempered distributions are canonical and continuously
injected into  Ultradistributions of Exponential Type
and as a consequence the Rigged
Hilbert Space  with tempered distributions is  canonical and continuously
included
in the Rigged Hilbert Space with  Ultradistributions of 
Exponential Type.

Another interesting property is that the space of UET is 
reflexive under the operation of Fourier transform (in a similar way
of tempered distributions of Schwartz)

In this paper we show that Ultradistributions of Exponential type
provides an adecuate framework 
for a consistent treatment of 
string and string field theories. In particular, a general
state of the closed bosonic bradyonic string is 
represented by UET of compact support,
and as a consequence the string field of a bradyonic 
bosonic string is a linear combination
of UET of compact support (CUET).

This paper is organized as follows:
in sections 2 and 3 we define the Ultradistributions of Exponential Type
and their Fourier transform. They
are  part of a Guelfand's Triplet ( or Rigged Hilbert Space \cite{tp12} )
together with their respective duals and a ``middle term'' Hilbert
space. 
In section 4 we treate the question of the equivalence
of Nanbu-Goto Lagrangian with a new Lagrangian for
the closed bradyonic bosonic string.
In section 5 we obtain  a  expression for the
Lagrangian of a closed tachyonic bosonic string.
In section 6 we give a expression for the propagator
of the string (do not confuse with the string field propagator 
of section 9).
In section 7 we give a new representation for the states 
of the string using CUET of compact support.
In section 8 we give  expressions for the field of the string,
the string field propagator and the creation and anihilation
operators of a string .
In section 9, we give expressions for the non-local action of a free string
and a non-local interaction lagrangian for the string field similar 
to  $\lambda{\phi}^4$ in Quantum Field Theory.
Also we show how to evaluate the convolution
of two string field propagators.
Finally, section 10 is reserved for a discussion of the principal results.

\section{Ultradistributions of Exponential Type}

Let ${\cal S}$ be the Schwartz space of rapidly decreasing test functions. 
Let ${\Lambda}_j$ be the region of the complex plane defined as:
\begin{equation}
\label{er2.1}
{\Lambda}_j=\left\{z\in\boldsymbol{\mathbb{C}} :
|\Im(z)|< j : j\in\boldsymbol{\mathbb{N}}\right\}
\end{equation}
According to ref.\cite{tp6,tp8} be the space of test functions $\hat{\phi}\in
{\large{V}}_j$ is
constituted by all entire analytic functions of ${\cal S}$ for which
\begin{equation}
\label{ep2.2}
||\hat{\phi} ||_j=\max_{k\leq j}\left\{\sup_{z\in{\Lambda}_j}\left[e^{(j|\Re (z)|)}
|{\hat{\phi}}^{(k)}(z)|\right]\right\}
\end{equation}
is finite.\\
The space $\large{Z}$ is then defined as:
\begin{equation}
\label{er2.3}
\large{Z} =\bigcap_{j=0}^{\infty} {\large{V}}_j
\end{equation}
It is a complete countably normed space with the topology generated by
the system of semi-norms $\{||\cdot ||_j\}_{j\in \mathbb{N}}$.
The dual of $\large{Z}$, denoted by
$\large{B}$, is by definition the space of ultradistributions of exponential
type (ref.\cite{tp6,tp8}).
Let $S$ the space of rapidly decreasing sequences. According to
ref.\cite{tp12} $S$ is a nuclear space. We consider now the space of
sequences $P$ generated by the Taylor development of
$\hat{\phi}\in\large{Z}$
\begin{equation}
\label{er2.4}
P=\left\{Q : Q
\left(\hat{\phi}(0),{\hat{\phi}}^{'}(0),\frac {{\hat{\phi}}^{''}(0)} {2},...,
\frac {{\hat{\phi}}^{(n)}(0)} {n!},...\right) : \hat{\phi}\in Z\right\}
\end{equation}
The norms that define the topology of $P$ are given by:
\begin{equation}
\label{er2.5}
||\hat{\phi} ||^{'}_p=\sup_n \frac {n^p} {n} |{\hat{\phi}}^n(0)|
\end{equation}
$P$ is a subespace of $S$ and therefore is a nuclear space.
As the norms $||\cdot ||_j$ and $||\cdot ||^{'}_p$ are equivalent, the correspondence
\begin{equation}
\label{er2.6}
{\large{Z}}\Longleftrightarrow P
\end{equation}
is an isomorphism and therefore $Z$ is a countably normed nuclear space.
We can define now the set of scalar products
\[<\hat{\phi}(z),\hat{\psi}(z)>_n=\sum\limits_{q=0}^n\int\limits_{-\infty}^{\infty}e^{2n|z|}
\overline{{\hat{\phi}}^{(q)}}(z){\hat{\psi}}^{(q)}(z)\;dz=\]
\begin{equation}
\label{er2.7}
\sum\limits_{q=0}^n\int\limits_{-\infty}^{\infty}e^{2n|x|}
\overline{{\hat{\phi}}^{(q)}}(x){\hat{\psi}}^{(q)}(x)\;dx
\end{equation}
This scalar product induces the norm
\begin{equation}
\label{er2.8}
||\hat{\phi}||_n^{''}=[<\hat{\phi}(x),\hat{\phi}(x)>_n]^{\frac {1} {2}}
\end{equation}
The norms $||\cdot ||_j$ and $||\cdot ||^{''}_n$ are equivalent, and therefore
$\large{Z}$ is a countably hilbertian nuclear space.
Thus, if we call now ${\large{Z}}_p$ the completion of
$\large{Z}$ by the norm $p$ given in (\ref{er2.8}), we have:
\begin{equation}
\label{er2.9}
\large{Z}=\bigcap_{p=0}^{\infty}{\large{Z}}_p
\end{equation}
where
\begin{equation}
\label{er2.10}
{\large{Z}}_0=\boldsymbol{H}
\end{equation}
is the Hilbert space of square integrable functions.\\
As a consequence the ``nested space''
\begin{equation}
\label{er2.11}
\Large{U}=\boldsymbol{(}\large{Z},
\boldsymbol{H},\large{B}\boldsymbol{)}
\end{equation}
is a Guelfand's triplet (or a Rigged Hilbert space=RHS. See ref.\cite{tp12}).

Any Guelfand's triplet
$\Large{G}=\boldsymbol{(}\boldsymbol{\Phi},
\boldsymbol{H},\boldsymbol{{\Phi}^{'}}\boldsymbol{)}$
has the fundamental property that a linear and symmetric operator
on $\boldsymbol{\Phi}$, admitting an extension to a self-adjoint
operator in
$\boldsymbol{H}$, has a complete set of generalized eigen-functions
in $\boldsymbol{{\Phi}^{'}}$ with real eigenvalues.

$\large{B}$ can also be characterized in the following way
( refs.\cite{tp6},\cite{tp8} ): let ${E}_{\omega}$ be the space of
all functions $\hat{F}(z)$ such that:

${\Large {\boldsymbol{I}}}$-
$\hat{F}(z)$ is analytic for $\{z\in \boldsymbol{\mathbb{C}} :
|Im(z)|>p\}$.

${\Large {\boldsymbol{II}}}$-
$\hat{F}(z)e^{-p|\Re(z)|}/z^p$ is bounded continuous  in
$\{z\in \boldsymbol{\mathbb{C}} :|Im(z)|\geqq p\}$,
where $p=0,1,2,...$ depends on $\hat{F}(z)$.

Let $N$ be:
$N=\{\hat{F}(z)\in {E}_{\omega} :\hat{F}(z)\; \rm{is\; entire\; analytic}\}$.
Then $\large{B}$ is the quotient space:

${\Large {\boldsymbol{III}}}$-
$\large{B}={E}_{\omega}/{N}$

Due to these properties it is possible to represent any ultradistribution
as ( ref.\cite{tp6,tp8} ):
\begin{equation}
\label{er2.12}
\hat{F}(\hat{\phi})=<\hat{F}(z), \hat{\phi}(z)>=\oint\limits_{\Gamma} \hat{F}(z) \hat{\phi}(z)\;dz
\end{equation}
where the path ${\Gamma}_j$ runs parallel to the real axis from
$-\infty$ to $\infty$ for $Im(z)>\zeta$, $\zeta>p$ and back from
$\infty$ to $-\infty$ for $Im(z)<-\zeta$, $-\zeta<-p$.
( $\Gamma$ surrounds all the singularities of $\hat{F}(z)$ ).

Formula (\ref{er2.12}) will be our fundamental representation for a tempered
ultradistribution. Sometimes use will be made of ``Dirac formula''
for exponential ultradistributions ( ref.\cite{tp6} ):
\begin{equation}
\label{er2.13}
\hat{F}(z)\equiv\frac {1} {2\pi i}\int\limits_{-\infty}^{\infty}
\frac {\hat{f}(t)} {t-z}\;dt\equiv
\frac {\cosh(\lambda z)} {2\pi i}\int\limits_{-\infty}^{\infty}
\frac {\hat{f}(t)} {(t-z)\cosh(\lambda t)}\;dt
\end{equation}
where the ``density'' $\hat{f}(t)$ is such that
\begin{equation}
\label{er2.14}
\oint\limits_{\Gamma} \hat{F}(z) \hat{\phi}(z)\;dz =
\int\limits_{-\infty}^{\infty} \hat{f}(t) \hat{\phi}(t)\;dt
\end{equation}
(\ref{er2.13}) should be used carefully.
While $\hat{F}(z)$ is analytic on $\Gamma$, the density $\hat{f}(t)$ is in
general singular, so that the r.h.s. of (\ref{er2.14}) should be interpreted
in the sense of distribution theory.

Another important property of the analytic representation is the fact
that on $\Gamma$, $\hat{F}(z)$ is bounded by a exponential and a power of $z$
( ref.\cite{tp6,tp8} ):
\begin{equation}
\label{er2.15}
|\hat{F}(z)|\leq C|z|^pe^{p|\Re(z)|}
\end{equation}
where $C$ and $p$ depend on $\hat{F}$.

The representation (\ref{er2.12}) implies that the addition of any entire function
$\hat{G}(z)\in N$ to $\hat{F}(z)$ does not alter the ultradistribution:
\[\oint\limits_{\Gamma}\{\hat{F}(z)+\hat{G}(z)\}\hat{\phi}(z)\;dz=
\oint\limits_{\Gamma} \hat{F}(z)\hat{\phi}(z)\;dz+\oint\limits_{\Gamma}
\hat{G}(z)\hat{\phi}(z)\;dz\]
But:
\[\oint\limits_{\Gamma} \hat{G}(z)\hat{\phi}(z)\;dz=0\]
as $\hat{G}(z)\hat{\phi}(z)$ is entire analytic
( and rapidly decreasing ),
\begin{equation}
\label{er2.16}
\therefore \;\;\;\;\oint\limits_{\Gamma} \{\hat{F}(z)+\hat{G}(z)\}\hat{\phi}(z)\;dz=
\oint\limits_{\Gamma} \hat{F}(z)\hat{\phi}(z)\;dz
\end{equation}

Another very important property of $\large{B}$ is that
$\large{B}$ is reflexive under the Fourier transform:
\begin{equation}
\label{er2.17}
\large{B}={\cal F}_c\left\{\large{B}\right\}=
{\cal F}\left\{\large{B}\right\}
\end{equation}
where the complex Fourier transform $F(k)$ of $\hat{F}(z)\in\large{B}$
is given by:
\[F(k)=\Theta[\Im(k)]\int\limits_{{\Gamma}_+}\hat{F}(z)e^{ikz}\;dz-
\Theta[-\Im(k)]\int\limits_{{\Gamma}_{-}}\hat{F}(z)e^{ikz}\;dz=\]
\begin{equation}
\label{er2.18}
\Theta[\Im(k)]\int\limits_0^{\infty}\hat{f}(x)e^{ikx}\;dx-
\Theta[-\Im(k)]\int\limits_{-\infty}^0\hat{f}(x) e^{ikx}\;dx
\end{equation}
Here ${\Gamma}_+$ is the part of $\Gamma$ with $\Re(z)\geq 0$ and
${\Gamma}_{-}$ is the part of $\Gamma$ with $\Re(z)\leq 0$
Using (\ref{er2.18}) we can interpret Dirac's formula as:
\begin{equation}
\label{er2.19}
F(k)\equiv\frac {1} {2\pi i}\int\limits_{-\infty}^{\infty}
\frac {f(s)} {s-k}\; ds\equiv{\cal F}_c\left\{{\cal F}^{-1}\left\{f(s)\right\}\right\}
\end{equation}
The treatment for ultradistributions of exponential type defined on
${\boldsymbol{\mathbb{C}}}^n$ is similar to the case of one variable.
Thus
\begin{equation}
\label{er2.20}
{\Lambda}_j=\left\{z=(z_1, z_2,...,z_n)\in{\boldsymbol{\mathbb{C}}}^n :
|\Im(z_k)|\leq j\;\;\;1\leq k\leq n\right\}
\end{equation}
\begin{equation}
\label{er2.21}
||\hat{\phi} ||_j=\max_{k\leq j}\left\{\sup_{z\in{\Lambda}_j}\left[
e^{j\left[\sum\limits_{p=1}^n|\Re(z_p)|\right]}\left| D^{(k)}\hat{\phi}(z)\right|\right]\right\}
\end{equation}
where $D^{(k)}={\partial}^{(k_1)}{\partial}^{(k_2)}\cdot\cdot\cdot{\partial}^{(k_n)}\;\;\;\;
k=k_1+k_2+\cdot\cdot\cdot+k_n$

${\large{B}}^n$ is characterized as follows. Let
${E}^n_{\omega}$ be the space of all functions $\hat{F}(z)$ such that:

${\Large {\boldsymbol{I}}}^{'}$-
$\hat{F}(z)$ is analytic for $\{z\in \boldsymbol{{\mathbb{C}}^n} :
|Im(z_1)|>p, |Im(z_2)|>p,...,|Im(z_n)|>p\}$.

${\Large {\boldsymbol{II}}}^{'}$-
$\hat{F}(z)e^{-\left[p\sum\limits_{j=1}^n|\Re(z_j)|\right]}/z^p$
is bounded continuous  in
$\{z\in \boldsymbol{{\mathbb{C}}^n} :|Im(z_1)|\geqq p,|Im(z_2)|\geqq p,
...,|Im(z_n)|\geqq p\}$,
where $p=0,1,2,...$ depends on $\hat{F}(z)$.

Let ${N}^n$ be:
${N}^n=\left\{\hat{F}(z)\in {E}^n_{\omega} :\hat{F}(z)\;\right.$
is entire analytic at minus in one of the variables $\left. z_j\;\;\;1\leq j\leq n\right\}$
Then ${\large{B}}^n$ is the quotient space:

${\Large {\boldsymbol{III}}}^{'}$-
${\large{B}}^n={E}^n_{\omega}/{N}^n$
We have now
\begin{equation}
\label{er2.22}
\hat{F}(\hat{\phi})=<\hat{F}(z), \hat{\phi}(z)>=\oint\limits_{\Gamma} \hat{F}(z) \hat{\phi}(z)\;
dz_1\;dz_2\cdot\cdot\cdot dz_n
\end{equation}
$\Gamma={\Gamma}_1\cup{\Gamma}_2\cup ...{\Gamma}_n$
where the path ${\Gamma}_j$ runs parallel to the real axis from
$-\infty$ to $\infty$ for $Im(z_j)>\zeta$, $\zeta>p$ and back from
$\infty$ to $-\infty$ for $Im(z_j)<-\zeta$, $-\zeta<-p$.
(Again $\Gamma$ surrounds all the singularities of $\hat{F}(z)$ ).
The n-dimensional Dirac's formula is
\begin{equation}
\label{ep2.23}
\hat{F}(z)=\frac {1} {(2\pi i)^n}\int\limits_{-\infty}^{\infty}
\frac {\hat{f}(t)} {(t_1-z_1)(t_2-z_2)...(t_n-z_n)}\;dt_1\;dt_2\cdot\cdot\cdot dt_n
\end{equation}
where the ``density'' $\hat{f}(t)$ is such that
\begin{equation}
\label{ep2.24}
\oint\limits_{\Gamma} \hat{F}(z)\hat{\phi}(z)\;dz_1\;dz_2\cdot\cdot\cdot dz_n =
\int\limits_{-\infty}^{\infty} f(t) \hat{\phi}(t)\;dt_1\;dt_2\cdot\cdot\cdot dt_n
\end{equation}
and the modulus of $\hat{F}(z)$ is bounded by
\begin{equation}
\label{er2.25}
|\hat{F}(z)|\leq C|z|^p e^{\left[p\sum\limits_{j=1}^n|\Re(z_j)|\right]}
\end{equation}
where $C$ and $p$ depend on $\hat{F}$.

\section{The Case N$\rightarrow\infty$}

\setcounter{equation}{0}

When the number of variables of the argument of the Ultradistribution of 
Exponential type tends to infinity we define:
\begin{equation}
\label{ep3.1}
d\mu(x)=\frac {e^{-x^2}} {\sqrt{\pi}}dx
\end{equation}
Let $\hat{\phi}(x_1,x_2,...,x_n)$ be such that:
\begin{equation}
\label{ep3.2}
\idotsint\limits_{-\infty}^{\;\;\infty}|\hat{\phi}(x_1,x_2,...,x_n)|^2 d{\mu}_1d{\mu}_2...
d{\mu}_n<\infty
\end{equation}
where
\begin{equation}
\label{ep3.3}
d{\mu}_i=\frac {e^{-x_i^2}} {\sqrt{\pi}}dx_i
\end{equation}
Then by definition
$\hat{\phi}(x_1,x_2,...,x_n)\in L_2({\mathbb{R}}^n,\mu)$
and 
\begin{equation}
\label{ep3.4}
L_2({\mathbb{R}}^{\infty},\mu)=
\bigcup\limits_{n=1}^{\infty}L_2({\mathbb{R}}^n,\mu)
\end{equation}
Let $\hat{\psi}$ be givem by
\begin{equation}
\label{ep3.5}
\hat{\psi}(z_1,z_2,...,z_n)={\pi}^{n/4}\hat{\phi}(z_1,z_2,...,z_n)
e^{\frac {z_1^2+z_2^2+...+z_n^2} {2}}
\end{equation}
where $\hat{\phi}\in {\large{Z}}^n$(the corresponding 
n-dimensional of $\large{Z}$).\\
Then by definition $\hat{\psi}(z_1,z_2,...,z_n)\in \large{G}({\mathbb{C}}^n)$,
\begin{equation}
\label{ep3.6}
\large{G}({\mathbb{C}}^{\infty})=\bigcup\limits_{n=1}^{\infty}
\large{G}({\mathbb{C}}^n)
\end{equation}
and the dual $\large{G}^{'}({\mathbb{C}}^{\infty})$ given by
\begin{equation}
\label{ep3.7}
\large{G}^{'}({\mathbb{C}}^{\infty})=\bigcup\limits_{n=1}^{\infty}
\large{G}^{'}({\mathbb{C}}^n)
\end{equation}
is the space of Ultradistributions of Exponential type.\\
The analog to (\ref{er2.11}) in the infinite dimensional case is:
\begin{equation}
\label{ep3.8}
\Large{W}=\boldsymbol{(}\large{G}({\mathbb{C}}^{\infty}),
L_2({\mathbb{R}}^{\infty},\mu), 
\large{G}^{'}({\mathbb{C}}^{\infty})\boldsymbol{)}
\end{equation}
If we define:
\begin{equation}
\label{ep3.9}
{\cal F}:\large{G}({\mathbb{C}}^{\infty})\rightarrow
\large{G}({\mathbb{C}}^{\infty})
\end{equation}
via the Fourier transform:
\begin{equation}
\label{ep3.10}
{\cal F}:\large{G}({\mathbb{C}}^n)\rightarrow
\large{G}({\mathbb{C}}^n)
\end{equation}
given by:
\begin{equation}
\label{ep3.11}
{\cal F}\{\hat{\psi}\}(k)=
\int\limits_{-\infty}^{\infty}\hat{\psi}(z_1,z_2,...,z_n)
e^{ik\cdot z+\frac {k^2} {2}}d{\rho}_1d{\rho}_2...d{\rho}_n
\end{equation}
where
\begin{equation}
\label{ep3.12}
d\rho(z)=\frac {e^{-\frac {z^2} {2}}} {\sqrt{2\pi}}\;dz
\end{equation}
we conclude that
\begin{equation}
\label{ep3.13}
\large{G}^{'}({\mathbb{C}}^{\infty})=
{\cal F}_c\{\large{G}^{'}({\mathbb{C}}^{\infty})\}=
{\cal F}\{\large{G}^{'}({\mathbb{C}}^{\infty})\}
\end{equation}
where in the one-dimensional case
\begin{equation}
\label{ep3.14}
{\cal F}_c\{\hat{\psi}\}(k)=
\Theta[\Im(k)]\int\limits_{{\Gamma}_+}\hat{\psi}(z)e^{ikz+\frac {k^2} {2}}\;d\rho-
\Theta[-\Im(k)]\int\limits_{{\Gamma}_{-}}\hat{\psi}(z)e^{ikz+\frac {k^2} {2}}\;d\rho
\end{equation}

\section{The Constraints for a Bradyonic Bosonic String}

\setcounter{equation}{0}

As is known the Nambu-Goto Lagrangian for the bosonic string 
is given by (\cite{tp13})
\begin{equation}
\label{ep4.1}
{\cal L}_{NG}=T\sqrt{({\dot{X}}\cdot X^{'})^{2}-{\dot{X}}^{2}X^{'2}}
\end{equation}
where
\[X=X(\tau,\sigma)\;\;;\;\;
\dot{X}={\partial}_{\tau}X\;\;;\;\;X^{'}={\partial}_{\sigma}X\]
If we use the constraint
\begin{equation}
\label{ep4.2}
(\dot{X}-X^{'})^{2}=0
\end{equation}
we obtain:
\begin{equation}
\label{ep4.3}
{\dot{X}}^{4}+X^{'4}=4(\dot{X}\cdot X^{'})^{2}-2{\dot{X}}^{2}X^{'2}\geq 0
\end{equation}
On the other hand 
\begin{equation}
\label{ep4.4}
({\dot {X}}^{2}-X^{'2})^{2}={\dot{X}}^{4}+X^{'4}-2{\dot{X}}^{2}X^{'2}
\end{equation}
and from (\ref{ep4.3}) we have
\begin{equation}
\label{ep4.5}
4{\cal L}_{BS}^2=T^2({\dot {X}}^{2}-X^{'2})^{2}=
4T^2[(\dot{X}\cdot X^{'})^{2}-{\dot{X}}^{2}X^{'2}]=4
{\cal L}_{NG}^{2}\geq 0
\end{equation}
As a consequence of (\ref{ep4.5}): 
\begin{equation}
\label{ep4.6}
{\cal L}_{NG}=T\sqrt{(\dot{X}\cdot X^{'})^{2}-{\dot{X}}^{2}X^{'2}}=
\frac {T} {2} |{\dot{X}}^{2}-X^{'2}|={\cal L}_{BS}
\end{equation}
We then see that is sufficient to use only one constraint  to obtain the
Lagrangian for a bosonic string theory from the Namb\'u-Goto
Lagrangian. Another constraint from which  (\ref{ep4.6}) follows is
\begin{equation}
\label{ep4.7}
(\dot{X}+X^{'})^{2}=0
\end{equation}
Thus, the equations  for the bosonic string reduce to:
\begin{equation}
\label{ep4.8}
\begin{cases}
\ddot{X}-X^{''}=0\\
(\dot{X}+X^{'})^2=0\\
X_{\mu}(\tau,0)=X_{\mu}(\tau,\pi)=0
\end{cases}
\end{equation}
or equivalently
\begin{equation}
\label{ep4.9}
\begin{cases}
\ddot{X}-X^{''}=0\\
(\dot{X}-X^{'})^2=0\\
X_{\mu}(\tau,0)=X_{\mu}(\tau,\pi)=0
\end{cases}
\end{equation}
The Euler-Lagrange equations for (\ref{ep4.8}) and (\ref{ep4.9}) are
respectively:
\[4\delta({\dot{X}}^2-X^{'2})[(\dot{X}\cdot\ddot{X}-X^{'}\cdot{\dot{X}}^{'}){\dot{X}}_{\mu}-
(X^{'}\cdot {\dot{X}}^{'}-X^{'}\cdot X^{''})X^{'}_{\mu}]+\]
\begin{equation}
\label{ep4.10}
Sgn({\dot{X}}^2-X^{'2})({\ddot{X}}-X^{''})+\lambda(\ddot{X}+2{\dot{X}}^{'}+
 X^{''})=0
\end{equation}
\[4\delta({\dot{X}}^2-X^{'2})[(\dot{X}\cdot\ddot{X}-X^{'}\cdot {\dot{X}}^{'}){\dot{X}}_{\mu}-
(X^{'}\cdot {\dot{X}}^{'}-X^{'}\cdot X^{''})X^{'}_{\mu}]+\]
\begin{equation}
\label{ep4.11}
Sgn({\dot{X}}^2-X^{'2})({\ddot{X}}-X^{''})+\lambda(\ddot{X}-2{\dot{X}}^{'}+
 X^{''})=0
\end{equation}
where $\lambda$ is a Lagrange multiplier.\\
If we define a physical state of the string as:
\begin{equation}
\label{ep4.12}
p^2|\Phi>=0
\end{equation}
A solution of (\ref{ep4.8}) defined on physical states is:
\begin{equation}
\label{ep4.13}
\begin{cases}
X_{\mu}(\tau,\sigma)=x_{\mu} + l^2 p_{\mu} \tau+\frac {il} {2}
\sum\limits_{n=-\infty\;;\;n\neq 0}^{\infty}
\frac {a_n} {n} e^{-2in(\tau-\sigma)}\\
p^2|\Phi>=0
\end{cases}
\end{equation}
It  is immediate to prove that (\ref{ep4.13}) is solution 
of Nambu-Goto equations on physical states, 
(Nambu-Goto equations arise from Euler-Lagrange equations 
corresponding to the  Lagrangian
(\ref{ep4.1}), and it is easy to prove that the currently used  solution 
for the  closed string movement 
is not solution of Nambu-Goto equations 
due to the fact that Virasoro operators $L_n$ and ${\tilde{L}}_n$
does not anihilate the physical states for $n<0$) and moreover,
does not form a set of commuting operators.\\
In a similar way a solution of (\ref{ep4.9}) at the is:
\begin{equation}
\label{ep4.14}
\begin{cases}
X_{\mu}(\tau,\sigma)=x_{\mu} + l^2p_{\mu} \tau+\frac {il} {2}
\sum\limits_{n=-\infty\;;\; n\neq 0}^{\infty}
\frac {{\tilde{a}}_n} {n} e^{-2in(\tau+\sigma)}\\
p^2|\Phi>=0
\end{cases}
\end{equation}

\section{The constraints for a Tachyonic Bosonic String}

\setcounter{equation}{0}

The Nambu-Goto Lagrangian for the tachyonic bosonic string is given by
\begin{equation}
\label{ep5.1}
{\cal L}_{NG}=T
\sqrt{{\dot{X}}^2X^{'2}-(\dot{X}\cdot X^{'})^2}
\end{equation}
If we use the constraint 
\begin{equation}
\label{ep5.2}
(\dot{X}\pm iX^{'})^2=0
\end{equation}
we obtain
\begin{equation}
\label{ep5.3} 
{\dot{X}}^4 + X^{'4}=2{\dot{X}}^2X^{'2}-4(\dot{X}\cdot X^{'})^2\geq 0
\end{equation}
On the other hand 
\begin{equation}
\label{ep5.4}
({\dot {X}}^{2}+X^{'2})^{2}={\dot{X}}^{4}+X^{'4}+2{\dot{X}}^{2}X^{'2}\geq 0
\end{equation}
and from (\ref{ep5.3}) we have
\begin{equation}
\label{ep5.5}
4{\cal L}_{BS}^2=T^2({\dot {X}}^{2}+X^{'2})^{2}=
4T^2[{\dot{X}}^{2}X^{'2}-(\dot{X}\cdot X^{'})^{2}]=4
{\cal L}_{NG}^{2}\geq 0
\end{equation}
As a consequence of (\ref{ep5.5}): 
\begin{equation}
\label{ep5.6}
{\cal L}_{NG}=T\sqrt{{\dot{X}}^{2}X^{'2}-(\dot{X}\cdot X^{'})^{2}}=
\frac {T} {2} |{\dot{X}}^{2}+X^{'2}|={\cal L}_{BS}
\end{equation}

\section{The propagator of the closed bosonic string}

\setcounter{equation}{0}

We write  $X_{\mu}$ in (\ref{ep4.13}) as:
\begin{equation}
\label{ep6.1}
X_{\mu}(\tau,\sigma)=x_{\mu} + l^2p_{\mu}\tau + l
\int\limits_{-\infty}^{\infty}\left(\frac {a_{\mu}(k)} {\sqrt{2|k|}}
e^{-i(|k|\tau-k\sigma)}+
\frac {a_{\mu}^+(k)} {\sqrt{2|k|}}e^{i(|k|\tau-k\sigma)}\right)\;dk
\end{equation}
where:
\[a_{\mu}(k)=i\sum\limits_{n>0}a_{\mu n}\delta(k-2n)\]
\begin{equation}
\label{ep6.2}
a_{\mu}^+(k)=-i\sum\limits_{n>0}a_{\mu n}^+\delta(k-2n)
\end{equation}
with:
\begin{equation}
\label{ep6.3}
[a_{\mu m},a^+_{\nu n}]={\eta}_{\mu\nu}
{\delta}_{mn}
\end{equation}
and
\[[a_{\mu}(k),a^+_{\nu}(k^{'})]={\eta}_{\mu\nu}\delta(k-k^{'})
\sum\limits_{n>0}\delta(k-2n)=\]
\begin{equation}
\label{ep6.4}
\frac {{\eta}_{\mu\nu}} {4}\Theta(k)\delta(k-k^{'})
\sum\limits_{n=-\infty}^{\infty}e^{-in\pi k}
\end{equation}
With the usual definition
\[{\Delta}_{\mu\nu} (\tau-{\tau}^{'},\sigma-{\sigma}^{'})=
 <0|T[X_{\mu}(\tau,\sigma)X_{\nu}({\tau}^{'},{\sigma}^{'})]|0>\]
the propagator for the string is (do not confuse with the 
string field propagator of section 9):
\[{\Delta}_{\mu\nu} (\tau-{\tau}^{'},\sigma-{\sigma}^{'})=\]
\begin{equation}
\label{ep6.5}
{\eta}_{\mu\nu}\frac {l^2} {4}\sum\limits_{n>0}n^{-1}
e^{-2inSgn(\tau-{\tau}^{'})[(\tau-{\tau}^{'})-(\sigma-{\sigma}^{'})]}
Sgn(\tau-{\tau}^{'})
\end{equation}
For $X_{\mu}$ in (\ref{ep4.14}) we obtain:
\[{\Delta}_{\mu\nu} (\tau-{\tau}^{'},\sigma-{\sigma}^{'})=\]
\begin{equation}
\label{ep6.6}
{\eta}_{\mu\nu}\frac {l^2} {4}\sum\limits_{n>0}n^{-1}
e^{-2inSgn(\tau-{\tau}^{'})[(\tau-{\tau}^{'})+(\sigma-{\sigma}^{'})]}
Sgn(\tau-{\tau}^{'})
\end{equation}

\section{A representation of  the states of the closed bosonic string}

\setcounter{equation}{0}

\subsection*{The case n finite}

For an ultradistribution of exponential type, we can write:
\[G(k)=\oint\limits_{{\Gamma}_z}\left\{\Theta[\Im(k)]\Theta[\Re(z)]-
\Theta[-\Im(k)]\Theta[-\Re(z)]\right\}\hat{G}(z)
e^{ikz}\;dz\]
\begin{equation}
\label{ep7.1}
\hat{G}(z)=\oint\limits_{{\Gamma}_k}\left\{\Theta[\Im(z)]\Theta[-\Re(k)]-
\Theta[-\Im(z)]\Theta[\Re(k)]\right\}\hat{G}(z)
e^{-ikz}\;dk
\end{equation}
and 
\[G(\phi)=\oint\limits_{{\Gamma}_k}
G(k)\phi(k)\;dk=\]
\begin{equation}
\label{ep7.2}
\oint\limits_{{\Gamma}_k}\oint\limits_{{\Gamma}_z}
\left\{\Theta[\Im(k)]\Theta[\Re(z)]-
\Theta[-\Im(k)]\Theta[-\Re(z)]\right\}\hat{G}(z)
\phi(k)e^{ikz}\;dk\;dz=
\end{equation}
\begin{equation}
\label{ep7.3}
-i\oint\limits_{{\Gamma}_k}\oint\limits_{{\Gamma}^{'}_z}
\left\{\Theta[\Im(k)]\Theta[\Im(z)]-
\Theta[-\Im(k)]\Theta[-\Im(z)]\right\}\hat{G}(-iz)
\phi(k)e^{kz}\;dk\;dz
\end{equation}
where the path ${\Gamma}^{'}_z$ is the path ${\Gamma}_z$
rotated ninety degrees counterclockwise around the origin
of the complex plane.

If $F(z)$ is an UET of compact support we can define: 
\[<\hat{F}(z),\phi(z)>=\]
\begin{equation}
\label{ep7.4}
\oint\limits_{{\Gamma}_k}\oint\limits_{{\Gamma}^{'}_z}
\left\{\Theta[\Im(k)]\Theta[\Im(z)]-
\Theta[-\Im(k)]\Theta[-\Im(z)]\right\}\hat{F}(z)
\phi(k)e^{kz}\;dk\;dz
\end{equation}
then:
\[<{\hat{F}}^{'}(z),\phi(z)>=\]
\[\oint\limits_{{\Gamma}_k}\oint\limits_{{\Gamma}^{'}_z}
\left\{\Theta[\Im(k)]\Theta[\Im(z)]-
\Theta[-\Im(k)]\Theta[-\Im(z)]\right\}{\hat{F}}^{'}(z)
\phi(k)e^{kz}\;dk\;dz=\]
\[-\oint\limits_{{\Gamma}_k}\oint\limits_{{\Gamma}^{'}_z}
\left\{\Theta[\Im(k)]\Theta[\Im(z)]-
\Theta[-\Im(k)]\Theta[-\Im(z)]\right\}{\hat{F}}(z)
k\phi(k)e^{kz}\;dk\;dz=\]
\begin{equation}
\label{ep7.5}
<\hat{F}(z),-z\phi(z)>
\end{equation}
If we define:
\begin{equation}
\label{ep7.6}
a=-z\;\;\;;\;\;\;a^+=\frac {d} {dz}
\end{equation}
we have
\begin{equation}
\label{ep7.7}
[a,a^+]=1
\end{equation}
Thus we have a representation for creation and annihilation operators
of the states of the string. The vacuum state annihilated 
by $z_{\mu}$ is the UET $\delta(z_{\mu})$,
and the orthonormalized states obtained by sucessive application of 
$\frac {d} {dz_{\mu}}$ to $\delta(z_{\mu})$ are:
\begin{equation}
\label{ep7.8}
F_n(z_{\mu})=\frac {{\delta}^{(n)}(z_{\mu})} {\sqrt{n!}}
\end{equation}
On the real axis:
\begin{equation}
\label{ep7.9}
<\hat{F}(z),\phi(z)>=\int\limits_{-\infty}^{\infty}\int\limits_{-\infty}^{\infty}
\overline{\hat{f}}(x)\phi(k)e^{kx}\;dx\;dk
\end{equation}
where $\overline{\hat{f}}(x)$is given by Dirac's formula:
\begin{equation}
\label{ep7.10}
\hat{F}(z)=\frac {1} {2\pi i}\int\limits_{-\infty}^{\infty}
\frac {\overline{\hat{f}}(x)} {x-z}\;dx
\end{equation}

A general state of the string can be writen as:
\[\phi(x,\{z\})=[a_0(x)+a^{i_1}_{\mu_1}(x){\partial}^{\mu_1}_{i_1}+
a^{i_1 i_2}_{\mu_1\mu_2}(x){\partial}^{\mu_1}_{i_1}{\partial}^{\mu_2}_{i_2}
+...+...\]
\begin{equation}
\label{ep7.11}
+a^{i_1i_2...i_n}_{\mu_1\mu_2...\mu_n}(x){\partial}^{\mu_1}_{i_1}
{\partial}^{\mu_2}_{\i_2}...{\partial}^{\mu_n}_{i_n}+...+...]
\delta(\{z\})
\end{equation}
where $\{z\}$ denotes $(z_{1\mu},z_{2\mu},...,z_{n\mu},...,....)$, and 
$\phi$ is a UET of compact support in the set of variables $\{z\}$.
The functions
$a^{i_1i_2...i_n}_{\mu_1\mu_2...\mu_n}(x)$
are solutions of
\begin{equation}
\label{ep7.12}
\Box a^{i_1i_2...i_n}_{\mu_1\mu_2...\mu_n}(x)=0
\end{equation}

\subsection*{The case n$\rightarrow \infty$}

In this case
\[G(k)=\oint\limits_{{\Gamma}_z}\left\{\Theta[\Im(k)]\Theta[\Re(z)]-
\Theta[-\Im(k)]\Theta[-\Re(z)]\right\}\hat{G}(z)
e^{ikz+\frac {k^2} {2}-\frac {z^2} {2}}\;\frac {dz} {\sqrt{2\pi}}\]
\[\hat{G}(z)=\oint\limits_{{\Gamma}_k}\left\{\Theta[\Im(z)]\Theta[-\Re(k)]-
\Theta[-\Im(z)]\Theta[\Re(k)]\right\}\times\]
\begin{equation}
\label{ep7.13}
\hat{G}(z)
e^{-ikz+\frac {z^2} {2}-\frac {k^2} {2}}\;\frac {dk} {\sqrt{2\pi}}
\end{equation}
\[G(\phi)=\oint\limits_{{\Gamma}_k}
G(k)\phi(k)e^{-k^2}\;\frac {dk} {\sqrt{\pi}}=\]
\[\oint\limits_{{\Gamma}_k}\oint\limits_{{\Gamma}_z}
\left\{\Theta[\Im(k)]\Theta[\Re(z)]-
\Theta[-\Im(k)]\Theta[-\Re(z)]\right\}\times\]
\begin{equation}
\label{ep7.14}
\hat{G}(z)
\phi(k)e^{ikz-\frac {z^2} {2}-k^2}\;\frac {dk\;dz} {\sqrt{2}\;\pi}=
\end{equation}
\[-i\oint\limits_{{\Gamma}_k}\oint\limits_{{\Gamma}^{'}_z}
\left\{\Theta[\Im(k)]\Theta[\Im(z)]-
\Theta[-\Im(k)]\Theta[-\Im(z)]\right\}\times\]
\begin{equation}
\label{ep7.15}
\hat{G}(-iz)
\phi(k)e^{kz+\frac {z^2} {2}-k^2}\;\frac {dk\;dz} {\sqrt{2}\;\pi}
\end{equation}
If $F(z)$ is an CUET we can define: 
\[<\hat{F}(z),\phi(z)>=\]
\[\oint\limits_{{\Gamma}_k}\oint\limits_{{\Gamma}^{'}_z}
\left\{\Theta[\Im(k)]\Theta[\Im(z)]-
\Theta[-\Im(k)]\Theta[-\Im(z)]\right\}\times\]
\begin{equation}
\label{ep7.16}
[\hat{F}(z)e^{-\frac {3z^2} {2}}]
\phi(k)e^{kz+\frac {z^2} {2}-k^2}\;\frac {dk\;dz} {\sqrt{2}\;\pi}=
\end{equation}
\[\oint\limits_{{\Gamma}_k}\oint\limits_{{\Gamma}^{'}_z}
\left\{\Theta[\Im(k)]\Theta[\Im(z)]-
\Theta[-\Im(k)]\Theta[-\Im(z)]\right\}\times\]
\begin{equation}
\label{ep7.17}
\hat{F}(z)
\phi(k)e^{kz-z^2-k^2}\;\frac {dk\;dz} {\sqrt{2}\;\pi}=
\end{equation}
and then
\[<-2z\hat{F}(z)+{\hat{F}}^{'}(z),\phi(z)>=\]
\[\oint\limits_{{\Gamma}_k}\oint\limits_{{\Gamma}^{'}_z}
\left\{\Theta[\Im(k)]\Theta[\Im(z)]-
\Theta[-\Im(k)]\Theta[-\Im(z)]\right\}\times\]
\[[-2z\hat{F}(z)+{\hat{F}}^{'}(z)]
\phi(k)e^{kz-z^2-k^2}\;\frac {dk\;dz} {\sqrt{2}\;\pi}=\]
\[-\oint\limits_{{\Gamma}_k}\oint\limits_{{\Gamma}^{'}_z}
\left\{\Theta[\Im(k)]\Theta[\Im(z)]-
\Theta[-\Im(k)]\Theta[-\Im(z)]\right\}\times\]
\[\hat{F}(z)k
\phi(k)e^{kz-z^2-k^2}\;\frac {dk\;dz} {\sqrt{2}\;\pi}=\]
\begin{equation}
\label{ep7.18}
<\hat{F}(z),-z\phi(z)>
\end{equation}
If we define:
\begin{equation}
\label{ep7.19}
a=-z\;\;\;;\;\;\;a^+=-2z+\frac {d} {dz}
\end{equation}
we have
\begin{equation}
\label{ep7.20}
[a,a^+]=1
\end{equation}
The vacuum state annihilated by $a$ is $\delta(z)e^{z^2}$. The orthonormalized
states obtained by sucessive application of $a^+$ are:
\begin{equation}
\label{ep7.21}
{\hat{F}}_n(z)=2^{\frac {1} {4}}{\pi}^{\frac {1} {2}}
\frac {{\delta}^{(n)}(z)e^{z^2}} {\sqrt{n!}}
\end{equation}
On the real axis we have
\begin{equation}
\label{ep7.22}
<\hat{F}(z),\phi(z)>=\int\limits_{-\infty}^{\infty}\int\limits_{-\infty}^{\infty}
\overline{\hat{f}}(x)\phi(k)e^{kx-x^2-k^2}\;\frac {dx\;dk} {\sqrt{2}\;\pi}
\end{equation}
where $\overline{\hat{f}}(x)$is given by Dirac's formula:
\begin{equation}
\label{ep7.23}
\hat{F}(z)=\frac {1} {2\pi i}\int\limits_{-\infty}^{\infty}
\frac {\overline{\hat{f}}(x)} {x-z}\;dx
\end{equation}

\section{The String Field}

\setcounter{equation}{0}

According to (\ref{ep4.13}) and section 7 the equation for the string field
is given by
\begin{equation}
\label{ep8.1}
\Box\Phi(x,\{z\})=({\partial}^2_0-{\partial}^2_1-{\partial}^2_2-{\partial}^2_3)
\Phi(x,\{z\})=0
\end{equation}
where $\{z\}$ denotes $(z_{1\mu},z_{2\mu},...,z_{n\mu},...,....)$, and 
$\Phi$ is a CUET in the set of variables $\{z\}$.
Any UET of compact support can be writed as a development of
$\delta(\{z\})$ and its derivatives. Thus we have:
\[\Phi(x,\{z\})=[A_0(x)+A^{i_1}_{\mu_1}(x){\partial}^{\mu_1}_{i_1}+
A^{i_1 i_2}_{\mu_1\mu_2}(x){\partial}^{\mu_1}_{i_1}{\partial}^{\mu_2}_{i_2}
+...+...\]
\begin{equation}
\label{ep8.2}
+A^{i_1i_2...i_n}_{\mu_1\mu_2...\mu_n}(x){\partial}^{\mu_1}_{i_1}
{\partial}^{\mu_2}_{\i_2}...{\partial}^{\mu_n}_{i_n}+...+...]
\delta(\{z\})
\end{equation}
where the quantum fields 
$A^{i_1i_2...i_n}_{\mu_1\mu_2...\mu_n}(x)$
are solutions of
\begin{equation}
\label{ep8.3}
\Box A^{i_1i_2...i_n}_{\mu_1\mu_2...\mu_n}(x)=0
\end{equation}
The propagator of the string field can be exppresed in terms of the propagators 
of the component fields:
\[\Delta(x-x^{'},\{z\},\{z^{'}\})=[\Delta_0(x-x^{'})+\Delta^{i_1j_1}_{\mu_1\mu_2}
(x-x^{'})\partial_{i_1}^{\mu_1}\partial_{j_1}^{'\nu_1}+...+...+\]
\begin{equation}
\label{8.4}
\Delta_{i_1...i_nj_1...j_n}^{\mu_1...\mu_n\nu_1...\nu_n}(x-x^{'})
\partial_{\mu_1}^{i_1}...\partial_{\mu_n}^{i_n}\partial_{\nu_1}^{'j_1}...
\partial_{\nu_n}^{'j_n}+...+...]\delta(\{z\},\{z^{'}\})
\end{equation}
We define the operators of annihilation and creation of a string as:
\[a(k,\{z\})=[a_0(k)+a_{\mu_1}^{i_1}(k)\partial_{i_1}^{\mu_1}+...+...+\]
\begin{equation}
\label{ep8.5}
a_{\mu_1...\mu_n}^{i_1...i_n}\partial_{i_1}^{\mu_1}...\partial_{i_n}^{\mu_n}
+...+...]\delta(\{z\})
\end{equation}
\[a^+(k,\{z^{'}\})=[a^+_0(k)+a_{\nu_1}^{+j_1}(k)\partial_{j_1}^{'\nu_1}+...+...+\]
\begin{equation}
\label{ep8.6}
a_{\nu_1...\nu_n}^{+j_1...j_n}\partial_{j_1}^{'\nu_1}...\partial_{j_n}^{'\nu_n}
+...+...]\delta(\{z^{'}\})
\end{equation}
If we define
\begin{equation}
\label{ep8.7}
[a_{\mu_1...\mu_n}^{i_1...i_n}(k),a_{\nu_1..\nu_n}^{+j_1...j_n}(k^{'})]=
f_{\mu_1...\mu_n\nu_1...\nu_n}^{i_1...i_nj_1...j_n}(k)\delta(k-k^{'})
\end{equation}
the commutations relations are
\[a(k,\{z\}),a^+(k^{'},\{z^{'}\})]=\delta(k-k^{'})[f_0(k)+f_{\mu_1\nu_1}^{i_1j_1}(k)
{\partial}_{i_1}^{\mu_1}{\partial}_{j_1}^{\nu_1}+...+...\]
\begin{equation}
\label{ep8.8}
f_{\mu_1...\mu_n\nu_1...\nu_n}^{i_1...i_nj_1...j_n}(k)
{\partial}_{i_1}^{\mu_1}...{\partial}_{i_n}^{\mu_n}
{\partial}_{j_1}^{\nu_1}...{\partial}_{j_n}^{\nu_n}+...+...]
\delta(\{z\},\{z^{'}\})
\end{equation}

\section{The Action for the String Field}

\setcounter{equation}{0}

\subsection*{The case n finite}

The action for the free bosonic bradyonic closed string field is:
\begin{equation}
\label{ep9.1}
S_{free}=
\oint\limits_{\{\Gamma_1\}}\oint\limits_{\{\Gamma_2\}}
 \int\limits_{-\infty}^{\infty}
\partial_{\mu}\Phi(x,\{z_1\})e^{\{z_1\}{\cdot}\{z_2\}}
\partial^{\mu}\Phi(x,\{z_2\})\;dx\;d\{z_1\}\;d\{z_2\}
\end{equation}
A possible interaction is given by:
\[S_{int}=\lambda\;\oint\limits_{\{\Gamma_1\}}
\oint\limits_{\{\Gamma_2\}}
\oint\limits_{\{\Gamma_3\}}
\oint\limits_{\{\Gamma_4\}}
\int\limits_{-\infty}^{\infty}
\Phi(x,\{z_1\})e^{\{z_1\}{\cdot}\{z_2\}}
\Phi(x,\{z_2\})e^{\{z_2\}{\cdot}\{z_3\}}
\Phi(x,\{z_3\})\times\]
\begin{equation}
\label{ep9.2}
e^{\{z_3\}{\cdot}\{z_4\}}
\Phi(x,\{z_4\})\;
dx\;d\{z_1\}\;d\{z_2\}
\;d\{z_3\}\;d\{z_4\}
\end{equation}
Both, $S_{free}$ and $S_{int}$ are non-local as expected.

\subsection*{The case n$\rightarrow \infty$}

In this case:
\begin{equation}
\label{ep9.3}
[S_{free}=
\oint\limits_{\{\Gamma_1\}}\oint\limits_{\{\Gamma_2\}}
 \int\limits_{-\infty}^{\infty}
\partial_{\mu}\Phi(x,\{z_1\})
e^{\{z_1\}{\cdot}\{z_2\}}
\partial^{\mu}\Phi(x,\{z_2\})\;dx\;d\{\eta_1\}\;d\{\eta_2\}
\end{equation}
where
\begin{equation}
\label{ep9.4}
d\eta(z)=\frac {e^{-z^2}} {\sqrt{2}\;\pi}
\end{equation}
and
\[S_{int}=\lambda\;\oint\limits_{\{\Gamma_1\}}
\oint\limits_{\{\Gamma_2\}}
\oint\limits_{\{\Gamma_3\}}
\oint\limits_{\{\Gamma_4\}}
\int\limits_{-\infty}^{\infty}
\Phi(x,\{z_1\})e^{\{z_1\}{\cdot}\{z_2\}}
\Phi(x,\{z_2\})e^{\{z_2\}{\cdot}\{z_3\}}
\Phi(x,\{z_3\})\times\]
\begin{equation}
\label{ep9.5}
e^{\{z_3\}{\cdot}\{z_4\}}
\Phi(x,\{z_4\})\;
dx\;d\{\eta_1\}\;d\{\eta_2\}
\;d\{\eta_3\}\;d\{\eta_4\}
\end{equation}
The convolution
of two propagators of  the string field is:
\begin{equation}
\label{ep9.6}
\hat{\Delta}(k,\{z_1\},\{z_2\})\ast
\hat{\Delta}(k,\{z_3\},\{z_4\})
\end{equation}
where $\ast$ denotes the convolution
of Ultradistributions of Exponential Type  
on the $k$ variable only.
With the use of the result
\begin{equation}
\label{ep9.7}
\frac {1} {\rho}\ast\frac {1} {\rho}=-\pi^2\ln\rho
\end{equation}
($\rho=x_0^2+x_1^2+x_2^2+x_3^2$ in euclidean space)

and
\begin{equation}
\label{ep9.8}
\frac {1} {\rho\pm i0}\ast\frac {1} {\rho\pm i0}=
 \mp i\pi^2\ln(\rho\pm i0)
\end{equation}
($\rho=x_0^2-x_1^2-x_2^2-x_3^2$ in minkowskian space)

the convolution of two string field propagators is finite.

\section{Discussion}

We have decided to begin this paper, for the benefit
of the reader, with a summary of the main characteristics
of Ultradistributions of Exponential Type and their Fourier
transform.

We have shown that UET are appropriate for
the description 
in a consistent way string and string field theories.
By means of  a new Lagrangian for the closed string strictly equivalent
to Nambu-Goto Lagrangian we have obtained a movement 
equation for the field of the string and solve it with the use of CUET 
We shown that this string field 
is a linear superposition of CUET.
We evaluate the propagator for the string field,
and calculate the convolution of two of them, taking
into account that string field theory is a non-local theory
of  UET of an infinite number of complex variables,
For practical calculations and experimental results 
we have given expressions that
involve only a finite number of variables.

As a final remark we would like to point out that our formulae
for convolutions follow from  general definitions. They are not
regularized expresions

\end{document}